\documentclass[aps,prl,twocolumn,superscriptaddress,showpacs]{revtex4}

\usepackage{graphicx}

\begin{document}


\title{Phonon Density of States of LaFeAsO$_{1-x}$F$_x$}


\author{A. D. Christianson}
\author{M. D. Lumsden}
\affiliation{Oak Ridge National Laboratory, Oak Ridge, TN 37831, USA}
\author{O. Delaire}
\affiliation{California Institute of Technology, W. M. Keck Laboratory, Pasadena, California 91125, USA}
\author{M. B. Stone}
\author{D. L. Abernathy}
\author{M. A. McGuire}
\author{A. S. Sefat}
\author{R. Jin}
\author{B. C. Sales}
\author{D. Mandrus}
\affiliation{Oak Ridge National Laboratory, Oak Ridge, TN 37831, USA}
\author{E. D. Mun}
\author{P. C. Canfield}
\affiliation{Ames Laboratory and Department of Physics and
Astronomy, Iowa State University, Ames, Iowa 50011, USA}
\author{J. Y. Y. Lin }
\author{M. Lucas}
\author{M. Kresch}
\author{J. B. Keith}
\author{B. Fultz}
\affiliation{California Institute of Technology, W. M. Keck Laboratory, Pasadena, California 91125, USA}
\author{E. A. Goremychkin}
\affiliation{Argonne National Laboratory, Argonne, IL 60439, USA}
\affiliation{ISIS Facility, Rutherford Appleton Laboratory, Chilton, Didcot OX11 0QX, United Kingdom}
\author{R. J. McQueeney}
\affiliation{Ames Laboratory and Department of Physics and
Astronomy, Iowa State University, Ames, Iowa 50011, USA}

\begin{abstract}

 We have studied the phonon density of states (PDOS) in LaFeAsO$_{1-x}$F$_x$ with inelastic neutron scattering methods.  The PDOS of the parent compound (x=0) is very similar to the PDOS of samples optimally doped with fluorine to achieve the maximum T$_c$ (x$\sim0.1$).  Good agreement is found between the experimental PDOS and first-principle calculations with the exception of a small difference in Fe mode frequencies.  The PDOS reported here is not consistent with conventional electron-phonon mediated superconductivity.
\end{abstract}


\maketitle

The discovery of superconductivity at 28 K in fluorine doped LaFeAsO by Kamihara \textit{et al}. and subsequent reports of T$_c$s in excess of 50 K in the broader RFeAsO (R=La,Ce,Pr,Nd,Sm,and Gd) family has generated great interest in the condensed matter physics community.\cite{LaOFFeAsdis,magnature,sefat,SmOFFeAsdis,Nddisc,singhlda,takahashi_pressure,cruz_lafeaso,kotliar} Although the superconductivity in the RFeAsO family is widely believed to be unconventional, there is no general consensus as to the precise mechanism behind superconductivity(\textit{e.g.} Ref. \cite{mazin2} and references therein).  Even the symmetry of the superconducting gap is controversial with reports of both one gap \cite{Chen}, and two gaps \cite{Wang,Matano,Ding} and reports of line nodes \cite{Matano} as well as a fully gapped Fermi surface \cite{Hashimoto,Ding}.

The critical temperatures of the RFeAsO family are exceeded only by the high-T$_c$ cuprates, so it is natural to ask if both materials share common physics.  In the cuprates, while significant changes are observed in the phonon spectrum as a function of doping and temperature\cite{arai_hightc,mcqueeney,Renker} the consensus is that superconductivity does not arise solely from electron-phonon coupling. However, electron-phonon coupling does play a significant role in the formation of competing charge-ordered ground states that appear in proximity to superconductivity\cite{Reznik}.  To date there have been only limited studies of the phonon modes in the RFeAsO family \cite{dong,qiu,raman}.  Therefore, to provide further insight into the phonon behavior in RFeAsO, we have performed inelastic neutron scattering measurements on samples independently prepared by two groups of the parent compound (LaFeAsO) as well as samples optimally doped with fluorine to achieve the maximum T$_c$ (LaFeAsO$_{0.9}$F$_{0.1}$).  We find little difference between the phonon frequencies of the parent compound and the superconducting counterpart.  The experimental phonon density of states (PDOS) is in excellent agrement with the first-principle calculations of Singh \textit{et al.}\cite{singhlda}.  The main exception being the observation of several Fe modes at $\sim$10\% lower energy in the experimental PDOS than predicted by calculations.  The experimental PDOS determined here is not consistent with \textit{conventional} electron-phonon mediated superconductivity in fluorine-doped LaFeAsO.

\begin{figure}
\centering\includegraphics[width=0.95\columnwidth]{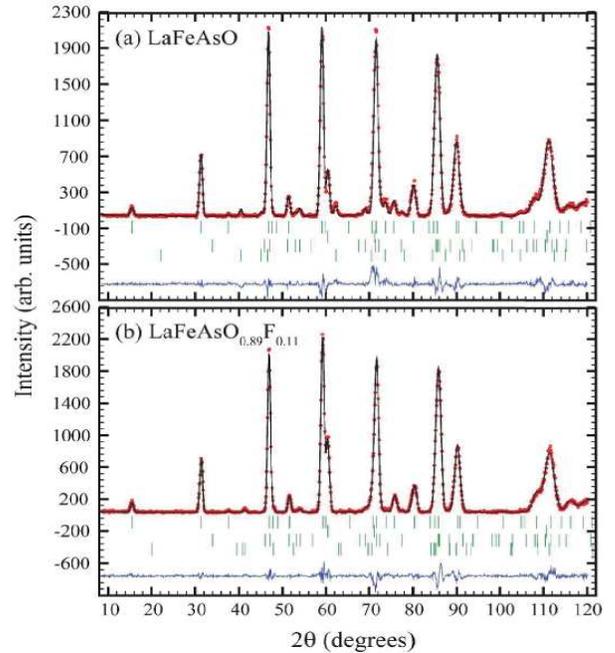}
\caption{\label{fig1} Neutron powder diffraction data: (a) LaFeAsO (S1) and (b) LaFeAsO$_{0.89}$F$_{0.11}$ (S2). In both panels the first three lines are the sample phase, aluminum sample can, and FeAs.  For (a) the fourth line is La$_2$O$_3$ while for (b) it is LaOF.}
\end{figure}

\begin{table}
\caption{\label{refineinfo} Sample composition determined from Reitveld refinement of neutron powder diffraction data.  The samples are denoted as follows: S1(S3) LaFeAsO and S2(S4) LaFeAsO$_{0.89}$F$_{0.11}$ (LaFeAsO$_{0.9}$F$_{0.1}$). Phase fractions are reported as weight percent.  The La$_2$O$_3$(LaOF) phases were found only in the parent (superconducting) samples.}
\begin{ruledtabular}
\begin{tabular}{ccccc}

   & S1 & S2 & S3 & S4 \\
   \hline
  Sample & 87.78 & 95.17 & 82.80 & 93.10 \\
  FeAs& 6.17 & 2.96 & 8.03 & 4.49 \\
  La$_2$O$_3$/(LaOF)& 6.05 & (1.87) & 9.17 & (2.42) \\

\end{tabular}
\end{ruledtabular}
\end{table}

Members of the RFeAsO family crystallize in the ZrCuSiAs structure type (space group P4/\textit{nmm}).  LaFeAsO has been found to undergo a structural phase transition at $\sim$160 K to a low temperature orthorhombic state followed by transition to an antiferromagnetic ordered state \cite{cruz_lafeaso,nomura_syncrotron,mcquire_big}.  Doping with fluorine suppresses the structural distortion and transition to long range magnetic order and superconductivity emerges with a maximum   T$_c$ $\sim$28 K  in LaFeAsO$_{0.89}$F$_{0.11}$ \cite{cruz_lafeaso,mcquire_big,qiu}.

We have investigated two sets of samples which were synthesized independently at Oak Ridge National Laboratory and at Ames Laboratory.  The Oak Ridge samples were synthesized by pelletizing a finely ground mixture of FeAs, La, La$_2$O$_3$, and LaF$_3$ (for the fluorine doped sample) in a glove box.  The resulting pellet was then sealed in a silica tube with 1/3 atm. argon and heated twice at 1200 $^{\circ}$C for 15 h.
All chemicals used were from Alfa Aesar and were of 99.95 \% purity or better.  The synthesis procedure resulted in samples of LaFeAsO (S1) and LaFeAsO$_{0.89}$F$_{0.11}$ (S2) of ~4.6 g each.  A T$_c$ of 27 K for S2 was determined by magnetic susceptibility measurements.  The Ames samples were synthesized as follows:  Polycrystalline LaFeAsO and LaFeAsO$_{0.9}$F$_{0.1}$ samples were synthesized by a standard solid reaction method. LaFeAsO samples were prepared using LaAs, Fe$_2$O$_3$ and Fe, while the 10\% F doped samples used LaAs, LaF$_3$, Fe$_2$O$_3$, Fe and As. A homogeneous mixture of starting materials was pressed into pellets and placed in a quartz tube, which was then sealed under Ar partial pressure. The entire preparation was done in N$_2$ filled glove boxes. The ampoule was kept at 600$^{\circ}$C for 5 hr before heating to 950$^{\circ}$C with ramp rate 50$^{\circ}$C/hr, held at temperature for 15 hr, and then annealed at 1150$^{\circ}$C for 50 hr. Finally, the ampoule was furnace-cooled.  The synthesis process resulted in samples of LaFeAsO (S3) and LaFeAsO$_{0.89}$F$_{0.11}$ (S4) of 5.0 g each.  The T$_c$ of 23 K for S4 was determined by magnetic susceptibility measurements.

For single-phonon scattering in a polyatomic material, the intensity of inelastic scattering of neutrons from a polycrystalline sample has the form,
\begin{equation}
\label{sqw}
S(Q,\omega)=\sum_{i} \sigma_i \frac{\hbar Q^2}{2 m_i} \exp(-2W_i)\frac{G_i(\omega)}{\omega}[n(\omega)+1],
\end{equation}
where $\sigma_i$ and $m_i$ are the neutron scattering cross section and mass of atom $i$, $\exp(-2W_i)$ is the Debye-Waller factor, $n(\omega)+1$ is the Bose occupation factor. $G_i(\omega)$ is defined as
\begin{equation}
\label{Gi}
G_i(\omega)=\frac{1}{3N} \sum_{j\textbf{q}} |\textbf{e$_i$}(j,\textbf{q})|^2 \delta[\omega-\omega(j,\bf{q})].
\end{equation}
where $\omega(j,\textbf{q})$ and $\textbf{e$_i$}(j,\textbf{q})$ are the frequencies and eigenvectors. Equations \ref{sqw} and \ref{Gi} show that inelastic neutron scattering measures the weighted PDOS.  In particular, the partial PDOS ($G_i(\omega)$) from each atomic species is weighted by $\sigma_i/m_i$.  For the elements in LaFeAsO, the relative magnitudes of this quantity are  0.23, 0.76, 0.27 and  1 for La, Fe, As, and O respectively and, hence, the vast majority of the scattered intensity is from Fe and O.  This weighting makes inelastic neutron scattering very sensitive to the presence of certain atoms and, consequently, understanding the levels of impurities are crucial.  To this end, we have performed neutron diffraction measurements at room temperature on all samples using the HB3 triple-axis spectrometer at the High Flux Isotope Reactor (HFIR) at Oak Ridge National Laboratory.  The results of these measurements are shown in Fig. 1 for S1 (a) and S2 (b).  Reitveld refinement of the diffraction data\cite{Carvajal-1993} indicates that in addition to the main phase, the samples contain small amounts of other phases as shown in Table 1\ref{refineinfo}.  A La(OH)$_3$ phase may also form in LaFeAsO samples, due to exposure in air.  This phase is of concern for the PDOS measurements as the $\sigma/m$ weighting makes this technique very sensitive to hydrogen in materials.  We only found the La(OH)$_3$ phase in S3 in very small quantities.



\begin{figure}
\centering\includegraphics[width=0.93\columnwidth]{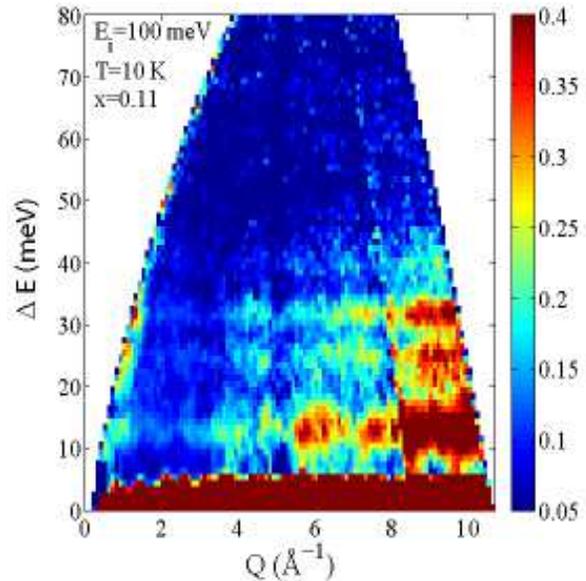}
\caption{\label{fig2} Inelastic neutron scattering intensity (color scale) of LaFeAsO$_{0.89}$F$_{0.11}$ (S2) at 10 K as a function of Q and E.  The scattering due to the empty sample holder has been removed.}
\end{figure}

The inelastic neutron scattering measurements reported here were performed on the ARCS direct geometry time-of-flight chopper spectrometer at the Spallation Neutron Source at Oak Ridge National Laboratory.  A closed cycle refrigerator provided sample temperatures from 10 to 300 K.  The data presented were measured with three different incident energies (E$_i$) / chopper frequencies, 130 meV / 600 Hz, 100 meV /  600 Hz, and 30 meV / 300 Hz.  These three settings result in an energy resolution of approximately 6.9 meV, 3.9 meV, and 1.2 meV respectively at the elastic position.  Limited inelastic scans at constant Q were performed on the HB3 triple-axis spectrometer at the HFIR (not shown) that provided independent confirmation of the ARCS data.

\begin{figure}
\centering\includegraphics[width=0.95\columnwidth]{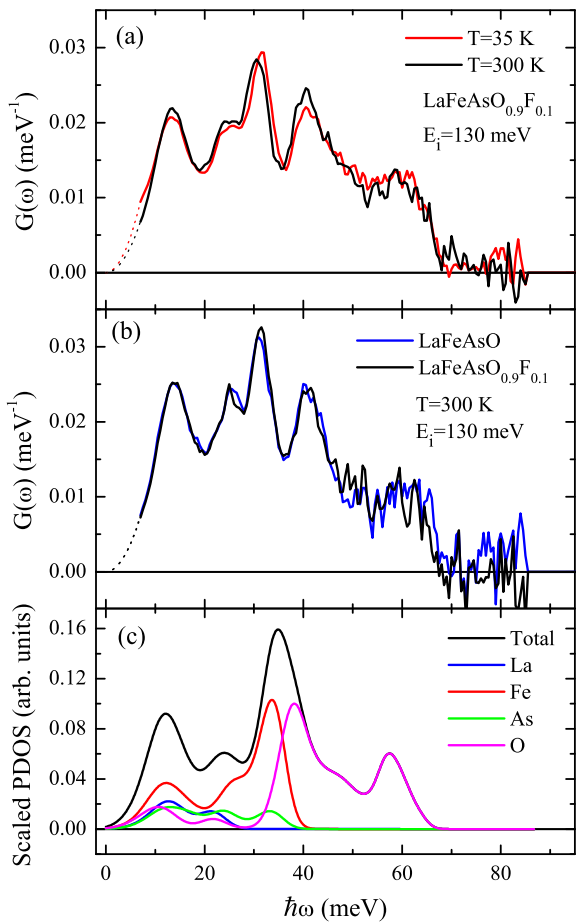}
\caption{\label{fig3} Phonon density of states of LaFeAsO$_{1-x}$F$_x$ (E$_i$=130 meV). (a) Experimental PDOS at 35 and 300 K for LaFeAsO$_{0.9}$F$_{0.1}$ (S4).  (b) Experimental PDOS for LaFeAsO (S3) and LaFeAsO$_{0.9}$F$_{0.1}$ (S4) at T=300 K. (c) First-principles calculations of Singh \textit{et al.}\cite{singhlda}.  The partial density of states have been normalized as described in the text.}
\end{figure}

The inelastic neutron scattering intensity of LaFeAsO$_{0.89}$F$_{0.11}$ (S2) as a function of Q and E at 10 K with an incident energy of 100 meV is shown in Fig.~\ref{fig2}. The scattering from the empty holder has been subtracted--no other corrections have been made to the data.  A number of excitations are clearly identifiable as lines of intensity at constant energy transfers with the strongest peaks observed near 12, 25, 30, and 40 meV.  The measured intensity becomes much stronger as Q increases allowing these modes to be identified as phonons due to the Q$^2$ dependence of the scattering law (Eq.~\ref{sqw}).  The intense line at zero energy transfer is the elastic line and in some cases the dispersion of acoustic modes is observed near Bragg points.  The extra intensity observed at low Q is due to low angle instrumental background.

\begin{figure}
\centering\includegraphics[width=0.95\columnwidth]{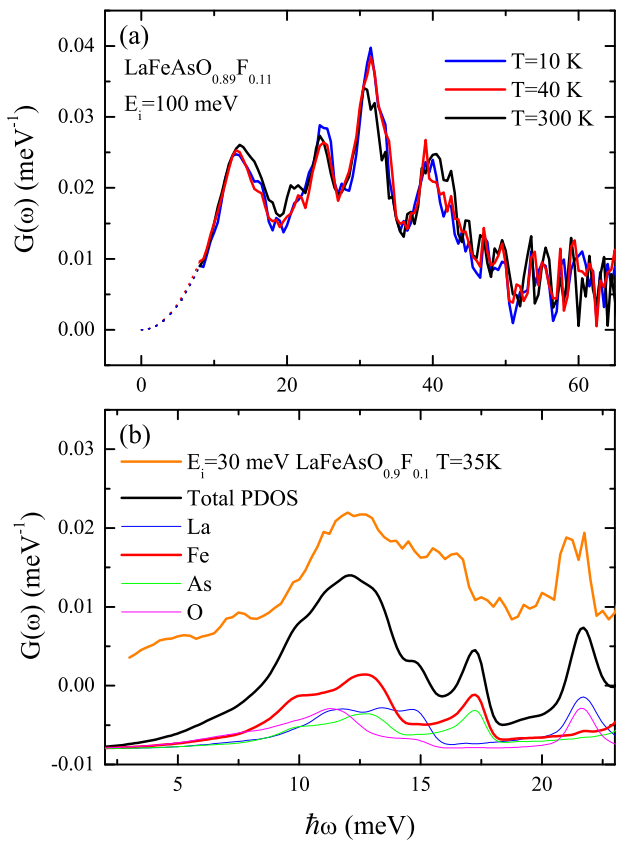}
\caption{\label{fig4} The Phonon Density of States of LaFeAsO$_{0.89}$F$_{0.11}$ (S2).  (a)Shows data with E$_i$=100 meV at 10, 40, and 300 K.  (b) Shows a comparison of high resolution data collected with E$_i$=30 meV at 300 K.}
\end{figure}

Figure \ref{fig3}(a) shows a comparison of LaFeAsO$_{0.9}$F$_{0.1}$ (S4) at 35 and 300 K.  The data were analyzed in the incoherent scattering approximation.  This analysis includes corrections for Bose and Debye-Waller factors, detector efficiency, as well as multiphonon scattering (including some account of multiple scattering), as described in \cite{drchops}.  The resulting PDOS is normalized to unity in all cases.  In Figs. \ref{fig3} and \ref{fig4} the dashed line indicates where the measured data ends and a quadratic approximation to the low energy PDOS begins.  Fig. \ref{fig3}(b) shows the PDOS at 300 K for the parent compound (S3) together with the superconductor (S4) for E$_i$ of 130 meV at 300 K.  There is no discernable difference between the inelastic response of the parent and that of the superconducting sample.  This is in stark contrast to the high T$_c$ cuprates where changes in oxygen concentration in YBa$_2$Cu$_3$O$_{7-\delta}$ leads to large changes in the PDOS at higher energies \cite{Renker,mcqueeney}.

In addition to the similarity between the parent and superconducting samples the phonon modes in the superconducting samples show only small variation with temperature.  The data of Fig. \ref{fig3}(a) at 35 K and 300 K are of high statistical quality, and are similar at nearly all energy transfers with both data sets clearly showing a phonon cutoff of 68 meV.  The major exception to this is a 1-2 meV stiffening of the 31 meV peak while cooling from 300 K to 35 K.  This stiffening is also shown in Fig. \ref{fig4}(a) which shows higher resolution measurements for the superconducting sample (S2) at temperatures of 300 K, 40 K, and 10 K.  Additionally, the peak at 12 meV softens slightly on cooling from 300 K to 40 K.  As shown in Fig. \ref{fig4}, no large changes are observed in cooling below T$_c$.



Figure \ref{fig3}(c) shows the results of first-principles calculations \cite{singhlda} of the bare PDOS of LaFeAsO. To allow for more direct comparison with measurements, the partial calculated PDOS is weighted by the relevant $\sigma/m$ ratio and convolved with an approximation of the instrumental resolution for E$_i$ of 130 meV.  In general, the calculation agrees very well with the measured PDOS.
The major deviation between experiment and theory is a number of the Fe modes which appear at systematically lower energies than the theoretical prediction.
This is most evident near 30 meV where calculations predicts an Fe mode near 34 meV while experiment shows a strong peak at 31 meV.  This represents a 10\% shift in the Fe mode energy.  To further emphasize this point, the Fig. \ref{fig4}(b) shows high resolution measurements taken with E$_i$ of 30 meV and the results of calculations treated in the manner described above.  The calculation predicts a mode at 17.2 meV which is not observed experimentally, but as in the case of the Fe mode at 31 meV a 10\% shift of the Fe mode frequencies in the calculation would explain the relatively larger density of states observed in the experimental PDOS near 15 meV.  This result is consistent with an inaccurately small Fe-As interatomic distance from local density approximation (LDA) calculations\cite{Yin,mazin2}, resulting in the prediction of higher phonon frequencies.

It is important to note that despite these deviations between experiment and theory, a 10\% shift in Fe mode frequencies would not appreciably change the Migdal-Eliashberg theory calculation of Boeri \textit{et. al} \cite{boeri}.  Alternatively, assuming that the experimental PDOS determined with inelastic neutron scattering approximates the electron-phonon spectral density, $\alpha^2F(\omega)$, (see for example Ref. \cite{Osborn_mgb2}) similar values are obtained to those found by Boeri \textit{et al.} \cite{boeri} for the phonon-electron coupling constant, logarithmic average of phonon frequency, and T$_c$ ($\sim$0.5 K).  As such, these measurements do not support conventional electron-phonon mediated superconductivity. We note that this analysis does not take into account non-standard electron phonon coupling mechanisms, such as the Fermi surface nesting effects known to be present in these compounds \cite{eschrig}.


In conclusion, we have measured the phonon density of states of the parent compound LaFeAsO and superconducting LaFeAsO$_{0.9}$F$_{0.1}$.  The measured phonon frequencies agrees well with first-principle calculations with the exception of a systematic shift of about 10\% of Fe mode frequencies.  Moreover these measurements are not consistent with conventional phonon-mediated superconductivity.

We thank D.J. Singh and M. H. Du for providing their calculations.  The portions of this work conducted at Oak Ridge National Laboratory were supported by the Scientific User Facilities Division and by the Division of Materials Sciences and Engineering, Office of Basic Energy Sciences, DOE.
Work at the Ames Laboratory was supported by the US Department of Energy – Basic Energy Sciences under Contract No. DE-AC02-07CH11358. This work was supported by the U.S. Department of Energy Office of Science, under Contract No. DE-AC02-06CH11357. This work benefitted from DANSE software developed under NSF award DMR-0520547.

\end{document}